\documentstyle[12pt,aps,preprint,prbbib]{revtex}
\newcommand{\be}{\begin{eqnarray}}
\newcommand{\ee}{\end{eqnarray}}
\newcommand{\la}{\langle}
\newcommand{\ra}{\rangle}
\newcommand{\p}{\partial}

\newcommand{\D}{V}
\newcommand{\cl}{{\cal L}}

\newcommand{\bmath}{\begin{mathletters}}
\newcommand{\emath}{\end{mathletters}}

\draft
\begin{document}

\preprint{ \scriptsize SUBMITTED TO BRIAN HEAD SPECIAL ISSUE OF JPC\hspace{4pt}}

\title{Electronic coherence in mixed-valence systems:\\ Spectral analysis}

\author{Younjoon Jung, Robert J. Silbey,
and Jianshu Cao\footnote{To whom correspondence should be addressed. Electronic mail: jianshu@mit.edu}}
\address{ Department of Chemistry, Massachusetts Institute of Technology \\
 Cambridge, MA 01239}

\date{\today}
\maketitle
\vspace{-0.5truein}

\begin{abstract}

The electron transfer kinetics of mixed-valence systems
is studied via solving the eigen-structure of the two-state non-adiabatic
diffusion operator for a wide range  of
electronic coupling constants  and energy bias constants.
The calculated spectral structure consists of three branches in 
the eigen-diagram, a real branch corresponding to exponential 
or multi-exponential decay and two symmetric branches
corresponding to population oscillations between donor and acceptor states.
The observed  electronic coherence is shown as a result of 
underdamped Rabi oscillations in an overdamped solvent environment.
The time-evolution of electron population is calculated by applying 
the propagator constructed from the eigen-solution to  the non-equilibrium 
initial preparation, and it agrees perfectly with the result of 
a direct numerical propagation of the density matrix. 
The resulting population  dynamics confirms that 
increasing the energy bias destroys electronic coherence.

\end{abstract}

\newpage

\section{INTRODUCTION}
Quantum coherence in  the dynamics of condensed phase systems
has become a subject of recent experimental and theoretical studies.
A central issue is the observability of electronic coherence 
in electron transfer systems given the fast dephasing time
in many-body quantum systems. 
Experimentally, with the advance in ultrafast laser technology,
oscillations in electronic dynamics  have been observed in photo-synthetic
reaction centers and other electron transfer systems
and are believed to arise from vibrational and/or electronic
coherence.\cite{vos-nat-93,jonas-jpc-95,arnett-jacs-95}
Accurate measurements on photo-induced electron transfer 
in mixed-valence compounds have demonstrated oscillations  
in  electronic populations on the femtosecond time-scale.\cite{vos-nat-93,reid-jpc-95}
Theoretically, detailed path-integral simulations 
suggest that such  oscillations take  place in 
electron transfer systems with large electronic coupling constants
and are sensitive to the initial preparation of the bath modes
 associated with the transfer processes.
Lucke \textit{et al.}\cite{lucke-jcp-97} extended the non-interacting blip approximation
to incorporate the non-equilibrium initial preparation 
and carried out extensive path-integral 
quantum dynamics simulations
for electron transfer reactions. According to their findings,  
large-amplitude oscillations are most likely to be observed 
in  symmetric mixed-valence systems
that are nearly adiabatic and with initial configurations
that are centered in the Landau-Zener crossing region.
Using the  transfer matrix technique,\cite{makarov-cpl-94}
Evans, Nitzan, and Ratner\cite{evans-jcp-98}
calculated short-time evolution  for the photo-induced
electron transfer reaction in ${\rm (NH_3)_5 Fe^{II}(CN)Ru^{III}(CN)_5}$.
Their results show fast oscillations in the electronic
population on the short time-scale(20 fs) followed by a slower population
relaxation on the long time-scale(100 fs). 
They  pointed out that these fast oscillations arise
 as the wave-function oscillates coherently between the donor and acceptor
states. The calculated long-time decay  rate is considerably smaller 
than the prediction by the golden-rule formulae,\cite{coalson-jcp-94,cho-jcp-95}
confirming the inadequacy of non-adiabatic rate theory
in studying mixed-valence systems.

In fact, a simple classical argument  helps understand the nature
of the observed oscillations.
As a function of  the ratio between $\lambda$ 
(the bath reorganization energy) and $\D$ (the electronic coupling constant),
there is a thermodynamic transition from the localized
electronic state in  a double-well potential 
to the delocalized electronic state in a single well 
potential.\cite{harris-jcp-83,silbey-jcp-84,carmeli-jcp-85,chandler-liquid-91,leggett-rmp-87}
(i) In the localized regime ($ \lambda \gg \D$),  
the large  reorganization energy destroys electronic coherence;
hence,  electron  transfer is an incoherent rate process,
which can be described  by  the non-interacting 
blip approximation or golden-rule rate in the non-adiabatic limit
and by  transition state  theory in the adiabatic limit.\cite{marcus-bba-85,newton-arpc-84,bader-jcp-90}
(ii) In the delocalized regime ($ \lambda\le \D $),  the electronic 
wave function  extends to both the donor and acceptor states
and electronic coherence persists over several oscillations.\cite{harris-jcp-83}
For  mixed-valence compounds, the electronic coupling constant 
is estimated to be in  the range of $10^3 \mbox{cm}^{-1}$, 
which is in the same order as the reorganization energy.\cite{vos-nat-93,evans-jcp-98}
Therefore, the observed oscillations and relaxation in mixed-valence 
systems are the consequence of a highly non-equilibrium 
coherence transfer process.

Due to  the delocalization nature of electronic states,
an adiabatic picture\cite{cao-cpl-99}
is more useful than the diabatic representation
for  analyzing the short-time dynamics in strongly-coupled systems.
In this picture, electronic coherence arises from Rabi oscillations
between two adiabatic surfaces and decays because of electronic dephasing.
Further, initial preparation and wave-packet dynamics can
modulate Rabi oscillations and  the overall  electronic dynamics.
Thus, the adiabatic representation provides a simple picture for
mixed-valence systems as well as a simple analytical method  
to model fast electron dynamics initiated by laser pulses.

As a general approach to describe condensed phase dynamics,
we recently proposed a spectral analysis method,\cite{cao-jcp-00} 
which is based on eigen-structures of dissipative 
systems instead of dynamic trajectories.
An important application of the approach is to analyze
a set of  two-state diffusion equations, 
which was first used by Zusman to treat solvent effects on electron transfer  
in the non-adiabatic limit.
The analysis allows us to  characterize 
multiple time-scales in  electron transfer processes 
including  vibrational relaxation, electronic coherence, 
activated curve crossing or barrier crossing. 
With this unified approach, the observed rate behavior, bi-exponential and
multi-exponential decay, and population 
oscillations are different components of the same kinetic spectrum.
Thus, several existing theoretical models,  developed for 
limited cases of electron transfer, can be analyzed, tested, and extended.
In particular, rate constants extracted from the analysis bridge smoothly
between the adiabatic and non-adiabatic limits, and the kinetic
spectrum in the large coupling regime reveals the nature of 
the localization-delocalization  transition as the consequence of 
two competing mechanisms.

In this paper,  the spectral analysis approach developed 
in Ref.\onlinecite{cao-jcp-00} is employed to  study 
the electron transfer dynamics  in mixed-valence systems.
We invoke the non-adiabatic diffusion equation proposed by Zusman 
to describe the electron transfer process in the over-damped 
solvent regime. 
As discussed earlier, electron transfer in mixed-valence systems takes place
in a different kinetic regime from the thermal activated regime described 
by Marcus theory. Thus, the time-scale separation is not satisfied, and
multi-exponential decay and oscillations are 
intrinsic nature of electron transfer kinetics. 
As a result, the kinetic spectra exhibit 
 bifurcation,  coalescence, and other complicated patterns.
Careful examination of these patterns
reveals the underlying mechanisms in mixed-valence systems.

The rest of the paper is organized as follows: 
The spectral structure of the non-adiabatic diffusion equation is 
formulated in Sec.~II.
Numerical examples of the spectral structure of  
strongly mixed electron transfer systems are presented and discussed
in Sec.~III and  concluding remarks are given in Sec.~IV.

\section{THEORY}
There have been extensive studies of the solvent effect on  
electron transfer dynamics in  literature with various 
 approaches.\cite{zusman-cp-80,calef-jpc-83,hynes-jpc-86,garg-jcp-85,sparpaglione-jcp-88}
One of the most extensively studied models for quantum dissipation 
is the spin-boson Hamiltonian,\cite{leggett-rmp-87,garg-jcp-85}
\be
H_{SB}={\epsilon \over 2}\sigma_{z}+V\sigma_{x}+
\sum_{\alpha}{p_{\alpha}^2\over 2m_{\alpha}}
+\sum_{\alpha}{1\over 2}m_{\alpha}\omega_{\alpha}^2
{\left(x_{\alpha}-\sigma_{z}
{c_{\alpha}\over{m_{\alpha}\omega_{\alpha}^2}}\right)}^2,
\label{sb}
\ee
where $\epsilon$ is  the energy bias between the two electronic  
states, $\D$ is the electronic coupling constant, 
 $\sigma_{z}$ and $\sigma_{x}$ are the usual Pauli matrices, and 
$\{x_{\alpha}$,$p_{\alpha}\}$ represents the bath degree of freedom 
 with mass $m_{\alpha}$, 
frequency $\omega_{\alpha}$, and the coupling constant $c_{\alpha}$. 
In this model effects of the bath modes on the dynamics of the system 
can be described via the spectral density defined by,
\be
J(\omega)={\pi\over2}\sum_{\alpha}{{c_{\alpha}^2}\over{m_{\alpha}\omega_{\alpha}}}\delta(\omega-\omega_{\alpha}). \label{spec}
\ee 

Equivalently, the  spin-boson Hamiltonian in Eq.~(\ref{sb})
can be separated into the electronic two-level part  $H_{TLS}$ and
the nuclear bath part  $H_{B}$,
\be
H_{SB}= H_{TLS} + H_B.
\label{two}
\ee
The two-level part of the Hamiltonian can be explicitly written as
\be
H_{TLS}(E)
=U_{1}(E)|1\ra\la 1|+U_{2}(E)|2\ra\la 2|+V(|1\ra\la 2|+|2\ra\la 1|),
\ee
where the diabatic energy surfaces $U_1(E)$ and $U_2(E)$ 
are functions of the stochastic variable $E$,
which  represents the polarization energy 
for a given solvent configuration.\cite{zusman-cp-80}
The transformation from the spin-boson Hamiltonian to the two-level 
system Hamiltonian has been shown in the literature\cite{garg-jcp-85,cao-jcp-97}
by the identity,
\be
E(\{x_{\alpha}\})=\sum_{\alpha}c_{\alpha}x_{\alpha}.
\ee
It is worthwhile to mention that the polarization energy $E$ was 
recognized as the reaction coordinate by Marcus in formulating  
non-adiabatic electron transfer theory.\cite{marcus-bba-85}
Since the electron transfer process involves the collective motion of
a large number of solvent degrees of freedom
and the two-level system is linearly coupled to the harmonic bath modes 
in the spin-boson Hamiltonian in Eq.~(\ref{sb}), 
the functional form for the free energy surface is harmonic,\cite{onuchic-jcp-93}
thus giving
\be
U_1(E)&=& {(E+\lambda)^2 \over {4 \lambda}},  \label{5a} \\
U_2(E)&=& {(E-\lambda)^2 \over {4 \lambda}} + \epsilon, \label{5b}
\ee
where $\lambda$ is the reorganization energy, which is related to the 
parameters in Eq.~(\ref{sb}),
\be
\lambda=\sum_{\alpha}{{c_{\alpha}^2}\over{2m_{\alpha}\omega_{\alpha}^2}} 
       ={1\over\pi}\int d\omega{J(\omega)\over\omega}. \label{lambda}
\ee   

Considering the fact that electron transfer processes 
are usually probed at room temperature in polar solvents,
we can treat the bath degrees of freedom in $H_{B}$  classically.
Then, the spin-boson Hamiltonian in Eq.~(\ref{two}) can
be used to derive a two-level classical equation of motion,
\be
i{\p \over {\p t}}\rho(t)=\cl\rho(t)=(\cl_B+ \cl_{TLS})\rho(t), \label{1}
\ee
where $i\cl_{B}=\{H_{B},\}$ is the  Poisson operator for the classical bath and
$\cl_{TLS}=[H_{TLS},\ ]/\hbar$ 
is  the Liouville operator for the two level system.
Explicitly, we express Eq.~(\ref{1}) in terms of the density matrix elements,
\bmath
\be 
\dot\rho_{1} &=& \cl_{1} \rho_{1} + i V (\rho_{12} - \rho_{21}), \label{2a} \\
\dot\rho_{2} &=& \cl_{2} \rho_{2} - i V (\rho_{12} - \rho_{21}), \label{2b} \\
\dot\rho_{12} &=& \cl_{12} \rho_{12} - i \omega_{12} \rho_{12} +
i V (\rho_{1}-\rho_{2}), \label{2c} \\
\dot\rho_{21} &=& \cl_{21} \rho_{21} + i \omega_{12} \rho_{21} - 
i V (\rho_{1}-\rho_{2}), \label{2d} 
\ee
\emath
where the Planck constant $\hbar$ is set to unity for simplicity,
$\rho_{i}$ is the diagonal matrix element for electronic population,
and $\rho_{ij}$ is the off-diagonal matrix element for electronic coherence.
Here, $\cl$ describes the relaxation process of classical bath,
with  $\cl_{i}$  defined on the free energy surface for the $i$th 
electronic state, and with $\cl_{12}$ and $\cl_{21}$ defined on the 
averaged free energy surface.
This set of semi-classical two-state equations  has
been previously derived in different context by several authors.
\cite{zusman-cp-80,garg-jcp-85,yang-jcp-89}  It should be mentioned that the mapping from 
the spin-boson Hamiltonian into the Zusman model requires the Lorentzian 
form of the spectral density,
\be
J(\omega)=2\lambda{{\omega\omega_c}\over{\omega^2+\omega_c^2}}. \label{lorentz}
\ee

Furthermore, we note that many chemically and biologically 
important electron transfer processes
take place in the over-damped solvent environment.
Therefore,  to describe the density matrix evolution in the 
electron transfer kinetics in the mixed-valence system, we 
invoke the non-adiabatic diffusion equation  proposed by Zusman.\cite{zusman-cp-80} 
Then, the bath relaxation operators in Eq.~(\ref{1}) are 
one-dimensional Fokker-Planck operators $\cl_{ij}$,
\be
&&\cl_{i} = D_E {\p \over \p E} \left ({\p \over \p E} + \beta {\p U_i(E) 
\over \p E} \right ), \label{3a} \\
&&\cl_{12} = \cl_{21} = {{\cl_{11}+\cl_{22}} \over 2}=D_E {\p \over \p E} 
\left ({\p \over \p E} + \beta {\p \bar{U}(E) \over \p E} \right) . \label{3b}
\ee
where $\beta={1/k_BT}$,
$\bar{U}$ and $\omega_{12}$ are the average  and the difference 
of the two free energy surfaces, respectively, 
\be
\bar{U}(E)={{U_1(E)+U_2(E)}\over{2}},\\ \label{4a}
\omega_{12}(E)=U_1(E)-U_2(E). \label{4b}
\ee
The energy diffusion constant $D_E$ is defined as
\be
D_E=\Omega\Delta_{E}^2, \label{7}
\ee
where  $\Delta_{E}^2$ is the mean square fluctuation 
of the solvent polarization energy
$$
\Delta_{E}^2={\la E^2\ra}=2\lambda k_BT, \label{6}
$$
and $\tau_D=1/\Omega$ is the 
the characteristic timescale of the Debye solvent.
The correlation function of the solvent polarization energy is given by
\be
C(t)=\la E(t)E(0)\ra=\Delta_E^2 \exp(-\Omega t).
\ee
Note that since the nuclear dynamics is modeled by the Fokker-Planck 
operator, the possibility of the vibrational coherence 
is excluded in this model of electron transfer dynamics.
It is  worthwhile to mention that one can obtain
the non-adiabatic diffusion equation starting from the spin-boson Hamiltonian,
by first deriving the evolution equation for the quantum dissipative dynamics,
 and then taking the semi-classical limit using the Wigner distribution 
functions, and finally assuming the over-damped diffusion limit.\cite{garg-jcp-85}

We investigate the spectral structure of the non-adiabatic diffusion 
operator by calculating the eigenvalues $\{-Z_{\nu}\}$ and the corresponding 
eigen-functions $\{|\psi_{\nu}\ra\}$. Hereafter we use  Greek indices to denote
the eigenstates and  Latin indices to denote the basis states of 
the non-adiabatic diffusion operator.
Because the non-adiabatic Liouville operator is non-Hermitian, 
the eigenvalues are generally given by complex values, and the right and left
eigen-functions corresponding to the same eigenvalue are not simply the
Hermitian conjugate to each other.\cite{simons-cp-73}
 For a given eigen-value $Z_{\nu}$,
the right and left eigen-functions of the non-adiabatic
diffusion operator are obtained from
\be
\cl |\psi_{\nu}^R \ra &=& -Z_{\nu}|\psi_{\nu}^R \ra,
\label{8a}\\
\la \psi_{\nu}^L|  \cl &=& -Z_{\nu}\la \psi_{\nu}^L |. 
\label{8b}
\ee
The method of eigenfunction solution is well known for the diffusion 
process on the harmonic potential energy surface.\cite{risken-fpe-84}
For a single quadratic potential $U(x)=\frac{1}{2} m\omega^2 x^2$,
the one-dimensional Fokker-Planck operator 
$\cl_{FP}=D({{\p^2}\over{\p x^2}}+{\beta {\p \over {\p x}} U'})$
can be transformed into the quantum mechanical Hamiltonian in imaginary time,
\be
H_s =-e^{\beta U(x)/2}\cl_{FP}e^{-\beta U(x)/ 2} 
    =-{1 \over {2\mu}} {\p ^2 \over {\p x^2}} + V_s(x), \label{11} 
\ee
where $\mu^{-1}=2D$, and the quadratic potential is 
\be
V_s(x) = D \left[{1 \over 4} (\beta U^{'}(x))^2 -{1 \over 2} \beta U^{''}(x) \right]
	= {1 \over 2} {\mu \gamma^2 x^2} - {\gamma \over 2}, \label{12}
\ee
with $\gamma=D m \omega^2/k_BT$. 
Since the transformed potential in Eq.~(\ref{12}) is just the same form as for
a simple harmonic oscillator with zero point energy compensation,
the eigenvalues and the eigen-functions for the original Fokker-Planck 
operator can be constructed immediately 
from the eigen-solutions of the  harmonic oscillator Hamiltonian.
Unlike the diffusion problem on the single potential energy surface, there
have been limited studies on the non-adiabatic diffusion problem involving 
more than one potential energy surface. 
 In this aspect, Cukier and co-workers have calculated 
the electron transfer rate by calculating the lowest eigenvalue of  
the non-adiabatic diffusion equation; however, 
their calculation was limited to the weak-coupling regime
where the Zusman rate is applicable.\cite{yang-jcp-89}

An important issue in solving  the non-adiabatic diffusion equation for
electron transfer is the choice of the basis functions
since  three different free energy surfaces are involved in Eq.~(\ref{1}):
two diabatic surfaces for the population density matrix elements and one 
averaged surface for the coherence density matrix element.
In this paper, the eigen-functions of $\cl_{12}$ are used as our basis set 
to represent the non-adiabatic diffusion equation. In principle, one could 
have chosen the eigen-functions of $\cl_{1}$ or $\cl_{2}$ as basis functions, 
however, in that case one has to evaluate appropriate Franck-Condon factors
when calculating the coupling matrix elements even with 
the Condon approximation.      
The Fokker-Planck operator $\cl_{12}$ is defined on  the averaged 
harmonic potential centered at $E=0$, and its eigen-solutions are
\be 
\cl_{12}| \phi^R_n \ra&=&-n \Omega | \phi^R_n \ra, \label{13a} \\ 
\la \phi^L_n|\cl_{12} &=&-n \Omega \la \phi^L_n |, \label{13b}
\ee
where  the right and left eigen functions are 
\be
\phi^R_n(E)={1\over{(2^n n!)^{1\over2}(2\pi\Delta_E^2)^{1\over4}}} 
\exp\left(-{{E^2}\over{2 \Delta_E^2}}\right) H_n\left({E\over {\sqrt{2} \Delta_E}}
\right), \label{14}
\ee
and
\be
\phi^L_n(E)={1\over{(2^n n!)^{1\over2}(2\pi\Delta_E^2)^{1\over4}}} 
H_n\left({E\over {\sqrt{2} \Delta_E}}\right), \label{15}
\ee
where $H_n$ is the $n$th order Hermite polynomial. 
As shown below, this choice of the basis set is convenient 
for our purpose.

To be consistent with the $\cl_{12}$ basis set, we 
separate the real and imaginary parts of the coherence density matrix, namely,
$u=$Re$\rho_{12}$ and $v=$Im$\rho_{12}$, and rewrite Eq.~(\ref{1}) as
\bmath
\be 
\dot\rho_{1} &=& (\cl_{12}+\delta\cl) \rho_1 - 2 V v, \label{9a} \\
\dot\rho_{2} &=& (\cl_{12}-\delta\cl) \rho_2 + 2 V v, \label{9b} \\
\dot u &=& \cl_{12} u + \omega_{12} v, \label{9c} \\
\dot v &=& \cl_{12} v - \omega_{12} u  + V (\rho_{1}-\rho_{2}), \label{9d} 
\ee
\emath
where we have defined $\delta\cl$ as 
\be
\delta\cl\ = {{\cl_{11}-\cl_{22}} \over 2}. \label{10}
\ee
Then, all the relevant operators in Eqs.~(\ref{9a})-(\ref{9d}) can be evaluated
 in terms of the right and left eigen-functions of $\cl_{12}$, giving
\be
\la \phi_{n}^{L} | \cl_{12}  | \phi_{m}^{R} \ra&=&-n\Omega\delta_{nm}, \label{16a}  \\
\la \phi_{n}^{L} | \delta\cl | \phi_{m}^{R} \ra&=&-\Omega\sqrt{{\lambda}
\over{2k_BT}}\sqrt{m+1}\delta_{n,m+1}, \label{16b}  \\
\la  \phi_{n}^{L} | \omega_{12} | \phi_{m}^{R} \ra&=&\sqrt{2\lambda k_BT}
(\sqrt{m}\delta_{n,m-1}+\sqrt{m+1} \delta_{n,m+1})-\epsilon\delta_{nm}, \label{16c} \\
\la  \phi_{n}^{L} | V| \phi_{m}^{R} \ra&=&V\delta_{nm}, \label{16d} 
\ee
where we  assume the  Condon approximation, i.e., the 
electronic coupling matrix element is independent
of the solvent degrees of freedom.
With the basis set, we can expand the density matrix elements as
\bmath
\be
\rho_{1}(E,t)&=&\sum_{n=0}^{\infty} a_{n}(t)\phi_{n}^{R}(E), \label{17a} \\
\rho_{2}(E,t)&=&\sum_{n=0}^{\infty} b_{n}(t)\phi_{n}^{R}(E), \label{17b} \\
u(E,t)&=&\sum_{n=0}^{\infty} c_{n}(t)\phi_{n}^{R}(E),  \label{17c} \\
v(E,t)&=&\sum_{n=0}^{\infty} d_{n}(t)\phi_{n}^{R}(E).  \label{17d}
\ee
\emath
Substituting Eqs.~(\ref{17a})-(\ref{17d})
into the eigenvalue equation Eq.~(\ref{8a}),
we have the following coupled linear equations 
\bmath
\be
-Z_{\nu} a_{n}&=&-n\Omega a_{n}-\Omega\sqrt{\lambda\over{2k_BT}}\sqrt{n}a_{n-1}
-2 V d_{n}, \label{18a} \\
-Z_{\nu} b_{n}&=&-n\Omega b_{n}+\Omega\sqrt{\lambda\over{2k_BT}}\sqrt{n}b_{n-1}
+2 V d_{n}, \label{18b} \\
-Z_{\nu} c_{n}&=&-n\Omega c_{n}+\sqrt{2\lambda k_BT}(\sqrt{n+1}d_{n+1}+\sqrt{n}d_{n-1})
-\epsilon d_{n}, \label{18c} \\
-Z_{\nu} d_{n}&=&-n\Omega d_{n}-\sqrt{2\lambda k_BT}(\sqrt{n+1}c_{n+1}+\sqrt{n}c_{n-1})+\epsilon c_{n} +V (a_{n}-b_{n}), \label{18d} 
\ee
\emath
which is an explicit basis set representation for 
the two-state diffusion operator in Eq.~(\ref{1}).
The linear equations for the left eigen-solution as defined by Eq.~(\ref{8b}) 
can be written by the transpose of Eqs.~(\ref{18a})-(\ref{18d}).
Diagonalizing the $4N\times 4N$ matrix ($N$ = number of basis functions)
defined in Eqs.~(\ref{18a})-(\ref{18d}), 
we obtain the eigenvalues $-Z_{\nu}$ and the corresponding eigenvectors 
of the non-adiabatic diffusion operator,
\be
|\psi_{\nu}^{R}\ra&=&\sum_{n} R_{n\nu}|\phi_{n}^{R}\ra, \label{eig1} \\
\la\psi_{\nu}^{L}|&=&\sum_{n} L_{\nu n}\la \phi_{n}^{L}|, \label{eig2} 
\ee
where $R_{n\nu}$ and $L_{\nu n}$ are  elements of the 
transformation matrices.

In general,
due to  the non-Hermitian nature of the non-adiabatic diffusion operator,
the right and left eigen-functions do not form an orthogonal set by themselves.
However, when the eigenvalues are all non-degenerate, 
the left and right eigen-functions form an orthogonal and complete set in  
dual Hilbert space.\cite{dahmen-condmat-99} 
Explicitly,  we have
\be
\sum_{n=0}L_{\nu n}R_{n\nu^{'}}=\delta_{\nu\nu^{'}},
\label{19}
\ee
for the orthogonality and
\be
\sum_{\nu} R_{n\nu}L_{\nu m}=\delta_{nm},
\label{20}
\ee
for the completeness. Using these properties,
we can construct the real time propagator for the operator $\cl$ as
\be
G(t)=\sum_{\nu}|\psi_{\nu}^{R}\ra \la\psi_{\nu}^{L}|
e^{-Z_{\nu}t}, \label{21}
\ee
and express the time evolution of the density matrix 
by  projecting a given initial distribution onto the eigenstates, giving
\be
|\rho(t)\ra=G(t)|\rho(0)\ra=\sum_{\nu}|\psi_{\nu}^{R}\ra 
\la\psi_{\nu}^{L}|\rho(0)\ra e^{-Z_{\nu}t}. \label{22}
\ee
Hence, the eigen-solution  to the two-state non-adiabatic diffusion equation
leads to  a complete description of  electron transfer dynamics.

\section{RESULTS AND DISCUSSIONS}

In the section, we present  the spectral
structure of the non-adiabatic diffusion operator  by diagonalizing
its matrix representation in Eqs.~(\ref{18a})-(\ref{18d}). 
In principle, we need infinite 
number of basis functions to diagonalize the non-adiabatic diffusion operator,
however, in practice, we have to truncate our basis set at some finite number. 
In all the calculations below,
we have used $N=50-200$ to diagonalize the $4N\times 4N$  matrix
and the effect of finite number basis on the spectral structure
has been  carefully examined.

\subsection{Spectral Structure}
\subsubsection{Mixed-valence systems}
In the mixed-valence compounds, the electronic coupling constant
has the same order of magnitude as the reorganization energy 
and  the electron transfer  dynamics
is usually probed  experimentally  at room temperature in  polar solvents. 
To study this process, Evans, Nitzan, and Ratner\cite{evans-jcp-98}  
carried out  real time path-integral simulations for the photo-induced
electron transfer reaction in ${\rm (NH_3)_5 Fe^{II}(CN)Ru^{III}(CN)_5}$.
Based on their model, 
we chose the parameters  for the calculation shown in Fig.~1 
as $\beta\Omega=0.6716$, $\beta\lambda=18.225$, $\beta\D=11.99$, and 
$\beta\epsilon=18.705$. As mentioned in the introduction
the mapping between the spin-boson Hamiltonian and 
the semi-classical Zusman equation is not rigorously defined. 
For example, for the non-adiabatic diffusion equation, the solvation energy
correlation function takes an exponential form with the rate $\Omega$,
whereas, for  the spin-boson model Hamiltonian,
it depends on the functional form of the spectral density.
It can be shown that the Ohmic spectral density with an exponential 
cut-off $\omega_{c}$ 
\be
J(\omega)=\eta\omega\exp(-\omega/\omega_{c}),
\ee
used in the calculation of Evans \textit{et al.}, leads to 
an energy correlation function with a Lorentzian form 
at high temperature,\cite{garg-jcp-85}
\be
C_{SB}(t)\approx {{2\eta\omega_{c} k_BT}\over{\pi}}{1\over{1+(\omega_{c}t)^2}}.
\ee
Then, the relaxation rate $\Omega$ used in our calculation 
is taken as the inverse of the mean  survival time of $C_{SB}(t)$,
 which is $\Omega=2\omega_{c}/{\pi}$. 

In Fig.~1 the spectral structure of the non-adiabatic operator is shown in  
complex space. We have used $N=200(4N=800)$ basis functions to calculate the 
eigenvalues. To remove the effect of finite basis set from the resulting 
spectral structure, 
we only show the first 400 eigenvalues in the complex plane. 
Since the non-adiabatic diffusion operator is non-Hermitian,
 the resulting spectrum shows complex conjugate paired eigenvalues 
as well as real eigenvalues, giving rise to  the tree structure 
 with three major branches (which we will call the \textit{eigen-tree}).
In Fig~.1, we  separate the real and imaginary parts of eigenvalue by
\be
-Z_{\nu}=-k_{\nu}-i\omega_{\nu}. \label{23}
\ee
Obviously, the real part, $k_{\nu}$, is always negative as all 
non-equilibrium physical quantities  decay to zero at time infinity,
 and it scales linearly with the index $\nu$ since the relaxation rate
corresponding to the $n$th basis state ${\phi_{n}}$ is proportional to $n$.
In general, the relative magnitudes of real and imaginary parts of 
eigenvalues determine  the time-evolution of the density matrix:
the  real eigenvalues correspond to the simple exponential decay components 
and the complex conjugate paired eigenvalues correspond to 
the damped oscillation components.

To classify the eigenvalues quantitatively 
according to their dynamic behavior, 
we introduce the dimensionless quantity $\theta_{\nu}$
\be
\theta_{\nu} \equiv { {2\pi k_{\nu}} \over {|{\omega_{\nu}}|} },  \label{24}
\ee
where $k_{\nu}$ is the decay rate and $2\pi/\omega_{\nu}$
is the oscillation period. The time-evolution of the density matrix 
component associated with the eigenvalue
 $Z_{\nu}$ is  an exponential decay if $\theta_{\nu} =\infty$,
an under-damped oscillation if  $\theta_{\nu} >1$,
 and a damped oscillation if $\theta_{\nu}\le 1$.
The relative amplitude of the each component
depends on the overlap matrix element between the initial 
density matrix and the eigenstate.
As an approximate criterion for the classification of the eigenvalues,
the slope corresponding to $\theta_{\nu}=1$ is shown 
in the eigen-tree diagram in Fig.~1.
There are a few eigenstates around and below the $\theta_{\nu}=1$ line, 
with a typical rate of $\beta k_{\nu}\approx 5$.
For the parameters used in the calculation, 
$\beta$ corresponds to $\sim$170 fs in real time, and,
therefore, these eigenstates exhibits damped oscillations with
a  period and a decay time in the femtosecond regime.
In their real-time path integral simulations,
 Evans \textit{et al.} showed that the population in the acceptor state 
oscillates with a few femtosecond period and these oscillation decays
in within 20 femtoseconds. 
Thus, qualitative features of the electron transfer dynamics 
can be predicted and understood
 from a careful examination of the spectral structure.
Since the spectral analysis presented here 
is based on the semi-classical diffusion equation while the 
path-integral study is based on the quantum mechanical spin-boson Hamiltonian,
the comparison between the two approaches
is expected to be qualitative.
In the following subsection,  further analysis 
 reveals the nature of these oscillations.

\subsubsection{Dependence on the coupling constant $\D$} 
To examine the underlying spectral structure in more details, 
eigenvalues of the non-adiabatic diffusion operator are plotted 
as functions of the electronic coupling constant in Fig.~2.
All the parameters except for the electronic coupling constant 
are the same as used in Fig.~1.

In Fig.~2(a), the real parts of the first 20  eigenvalues  are shown
as functions of the electronic coupling constant. 
Note that eigenvalues corresponding to complex conjugate pairs 
have the same real part, thus they coalesce in the real eigenvalue diagram.  
When the coupling constant is very small $(\beta\D \ll 1)$,
 the real part of the first non-zero eigenvalue is very well separated
from the eigenvalues of excited states, 
so the dynamics of  electron transfer can be considered as a
 incoherent rate process with a well-defined rate constant, $k_1$. 
When the coupling constant is larger $(\beta\D \approx 1)$,
the first excited state  becomes close to the second excited state, and 
they start to merge into a  complex conjugate pair. If the 
coupling constant increases further, eigen-values show a 
bifurcation behavior at $\beta\D \approx 10$. 
Therefore, in this regime, the electron transfer kinetics 
show multiple time-scale relaxation as well as  coherent oscillation.
The complicated behavior of 
coalescence and bifurcation in the real eigenvalue
appears more frequently at higher states.

Another interesting feature of the real eigenvalue diagram  
is that a set of  real eigenvalues decreases consistently 
as the coupling constant increases from zero.
It turns out that these eigenstates take on large imaginary parts, 
which are responsible for the onset 
of the imaginary branches of the eigen-tree. 
In Fig.~2(b), the imaginary parts of the lowest 30 eigenvalues are plotted 
as  functions of the coupling constant.   Interestingly,
the imaginary part of the eigenvalue increases approximately 
linearly with the coupling constant at large coupling  regime.
In fact, the dependence on the coupling constant is similar to that of
 the Rabi frequency for the two-level system,
\be 
\Omega_{R}=\sqrt{\epsilon^2+4\D^2},
\ee
which is shown  in Fig.~2(b).
As pointed out in a recent paper,\cite{cao-cpl-99}
electronic coherence in mixed-valence systems arises from
 Rabi oscillations between two adiabatic surfaces
and decays because of dephasing.

To demonstrate 
the correlation of the real and imaginary parts of the eigenvalues as 
functions of the coupling constant, we present a three dimensional plot of 
the spectral structure in Fig.~2(c). For clarity,
 only the positive branches of the imaginary eigenvalues are shown. 
If we compare Fig.~2(c) with  Fig.~2(a),
the very rapidly decaying states shown in Fig.~2(a) 
take on large imaginary parts corresponding to the Rabi oscillations
as the coupling constant increases, and 
these states are responsible for the onset 
of the imaginary branches in the eigen-tree for 
the mixed-valence system shown in Fig.~1. 

\subsection{Density Matrix Propagation}

To check the validity of the spectral analysis as a density matrix propagation 
scheme, we calculated the time-evolution of the density matrix by applying the 
propagator defined by Eq.~(\ref{21}) to the initial density matrix for various energy biases. 
Although it may seem straightforward to use the spectral method as a propagation scheme, the case for a non-Hermitian operator is not trivial and has not been explored. The main reason is that though the left and right 
eigen-functions of a non-Hermitian operator can be shown to form a 
bi-orthogonal set for the non-degenerate eigenvalue case,
numerically these eigen-functions may not be stable enough to be used as a 
complete orthonormal basis for the density matrix propagation, especially  
in the nearly degenerate eigenvalue case. 
We can understand the situation as follows: When the two nearly 
degenerate eigenvalues $Z_{1}$ and $Z_{2}$ are 
obtained from a non-Hermitian operator, 
the orthogonality implies that $\la L_{2}|$ and $|R_{1}\ra$ 
are orthogonal to each other as well as $\la L_{1}|$ and $|R_{2}\ra$. When two 
eigenvalues become very close to each other, unlike the Hermitian operator case, 
$\la L_{1}|$ and $\la L_{2}|$ almost coincide and so do $|R_{1}\ra$ and $|R_{2}\ra$, 
so that $\la L_{1}|$ and $|R_{1}\ra$ become almost orthogonal to each other. To still 
satisfy the normalization condition $\la L_{n}|R_{n}\ra=1$ in this case, 
the eigenfunction should be scaled up, thus making the spectral structure 
very sensitive to the numerical error involved in the calculation of eigenfunctions.  
For an interesting discussion on this point, 
one may refer to the work by Nelson and co-workers.\cite{dahmen-condmat-99} Due to this 
numerical instability, the use of the spectral method as a density matrix 
propagation scheme is not without limitation.

Figure 3a shows the spectral structure and the time-evolution of the 
density matrix propagation for the case of  
$\beta\Omega=1$, $\beta\lambda=15$, $\beta\D=12$, and $\beta\epsilon=5$. 
Generally, when the energy bias is small ($\beta\epsilon\le 5$), 
the left and right eigenfunctions can form a complete orthonormal basis set, 
so the spectral method is stable and can be used 
as a numerical propagation method for the density matrix. 
With a large energy bias, however, the calculated eigenfunctions 
may not form a complete orthonormal basis.
To model for the photo-induced back electron transfer
experiment in the mixed-valence compounds the initial density matrix is chosen  
as a thermal equilibrium distribution of the donor state(i.e. 1-state)
pumped to the acceptor state(i.e. 2-state),\cite{reid-jpc-95,lucke-jcp-97,evans-jcp-98}
\bmath
\be
\rho_{i}(E,0)&=&{1 \over {{\sqrt{2\pi}} \Delta_{E}}} \exp \left(-{{(E+\lambda)^2} 
\over {2\Delta_E^2}}\right)\delta_{i2}, \label{28} \\
\rho_{12}(E,0)&=&\rho_{21}(E,0)=0. \label{29}
\ee
\emath
It would be straightforward to calculate the spatial distribution of the density 
matrix in time $\rho(E,t)$ by applying the propagator in Eq.~(\ref{21}) to the
above initial density matrix; however, to demonstrate the overall temporal 
behavior only the time evolution 
of the total population in the acceptor state is calculated,
\be
P_{2}(t)=\int dE\rho_{2}(E,t). \label{30}
\ee
In order to check the validity of the spectral method as a propagation 
scheme in this case, we also calculated the time evolution of the density 
matrix by directly solving the $4N$ differential equations 
for the expansion coefficients of the 
density matrix using the Bulirsh-Stoer algorithm,\cite{press-numrecipes-92} and the comparison 
in Fig.~3(a) shows a perfect agreement. If only the transient behavior is 
concerned with, the direct propagation method would be preferred over the 
spectral method, however, the spectral propagation has the advantage when 
calculating the long time behavior once the complete spectrum is known. 
Overall, the computational costs for two method are comparable to each other. 
As expected from the spectral structure shown in the previous section 
the population in the acceptor state shows an underdamped 
coherent oscillation behavior at initial times 
followed by a damped oscillation behavior at later times.

Further, we have also studied the density matrix propagation for different energy biases to 
examine the electronic dephasing effect. As seen from Fig.~4(a), 
the increase in energy bias 
destroys the electronic coherence dramatically. Another interesting 
observation is the phase shift in the population dynamics 
as the energy bias is varied, 
and it is because the Rabi oscillation frequency increases with energy bias.
We can confirm the temporal behavior of the density matrix propagation by 
examining the spectral structure shown in Fig.~4(b). 
The period of the initial coherence is estimated to be $\tau_{osc}\approx 0.25\beta$ 
from Fig.~4(a). In comparison, the Rabi frequency for the corresponding adiabatic 
two-level system is given by 
$\Omega_{R}=\sqrt{\epsilon^2+4V^2}\approx 25{\beta}^{-1}$, 
which can also be obtained from the onset of imaginary 
branches in the eigen-tree shown in Fig.~4(b),   
and the estimation is consistent with the oscillation period observed in the dynamics 
since $\tau_{osc}\approx 2\pi/{\Omega_{R}}$. 
The real eigenvalues of the lowest excited states in the the imaginary branches are 
estimated to be $\beta k\approx 1-2$, and they agree with the decay time 
of the oscillation amplitude in Fig.~4(a), confirming the validity of the 
spectral method as a density matrix propagation scheme. 
Even though it has been well known in the literature that 
the damping of population is enhanced with 
increased energy asymmetry,\cite{leggett-rmp-87} we have also confirmed 
this through the spectral analysis method.
  
As an example of the eigenfunction responsible for the coherent 
oscillation behavior observed in Fig.~4 (b), we show 
the left and right eigenfunctions corresponding to a  
complex eigenvalue $\beta Z=2.6228\pm i 26.394$ for a symmetric case
$(\beta\epsilon=0)$ and $\beta Z=2.8057\pm i 26.466$ for 
an asymmetric case$(\beta\epsilon=5)$ in Figs.~5 and 6.  
The eigenfunctions corresponding to a complex conjugate pair 
of eigenvalues are also complex conjugate to each other; 
therefore, the frequency spectrum of the density matrix evolution is 
proportional to the norm of wavefunction. 
We note that the left eigenfunction is more extended than 
the right eigenfunction. Although the population distribution 
in the donor and acceptor states corresponding to coherent oscillation 
is inverted with respect to the Boltzmann distribution, 
it does not contribute to the steady-state population 
distribution due to the transient nature.

\section{CONCLUDING REMARKS}

In this paper we have applied the spectral analysis method to
the non-adiabatic two-state diffusion equation,
that describes  electron transfer dynamics in Debye solvents.
In particular, we have examined electronic coherence  
in  mixed-valence compounds, and  
demonstrated that underdamped Rabi oscillations 
are observed in an overdamped solvent environment.  
Detailed study of the spectral structure of the non-adiabatic operator 
for various  energy biases and coupling constants allows us
to determine the underlying mechanisms of electron transfer kinetics.
Eigenvalues form three branches in the eigen-diagram: a single branch of 
real eigenvalues and two symmetric branches of complex conjugate eigenvalues.
In strongly coupled systems, all three branches have a similar order of 
magnitude, indicating that both multiple-exponential decay and coherent 
oscillations can be observed experimentally.

We have investigated the dependence of the spectral structure on 
the coupling constant. In the very weak coupling regime, 
the lowest excited state is well separated from higher 
states, which makes the electron transfer dynamics a well-defined 
rate process. In the strong coupling regime, however, the eigenvalue 
diagram shows coalescence/bifurcation behavior in the complex plane. 
We have used the spectral method to calculate the time-evolution 
of the density matrix, and indeed, observed  
electronic coherence in the temporal 
behavior of population in the acceptor state for 
non-equilibrium initial distributions. 
We also found a good agreement between results of 
the spectral propagation method 
and of the numerical propagation method for small energy bias cases.
Due to non-Hermitianity of the non-adiabatic operator, the 
spectral propagation method was not numerically stable 
for large energy bias cases.

For an isolated quantum system,
the eigen-solution to the Schr\"odinger equation completely
determines its dynamics.  In a similar fashion,
the eigen-solution to the non-adiabatic diffusion operator 
completely characterizes the dynamics of a dissipative system
and thus provides a powerful tool to analyze dissipative dynamics. 
It is well known that quantum dynamics comes from the underlying 
spectra, especially in gas-phase chemical systems;\cite{field-acp-97}
however,  the spectral aspect of condensed phase dissipative systems 
has not been well recognized yet and  deserves further investigation.
Though the analysis presented here is restricted to  
semi-classical dissipative systems, it may also be applied to  quantum 
dissipative dynamics. In principle,
we can derive  the evolution equation for quantum 
dissipative systems either from first principles or 
through numerical reduction,
and then pose the quantum dissipative equation of motion
as an eigen-value problem. 
Along this line, the dissipative dynamics of the spin-boson Hamiltonian, 
which has been studied mostly as a dynamic problem,\cite{makarov-cpl-94,wang-jcp-110}
can also be explored  as a spectral problem in the future.

\section*{ACKNOWLEDGMENTS}
The authors would like to thank NSF for financial support.
One of us (YJ) would like to thank the Korean Foundations for Advanced Studies 
for financial support.

\newpage
\begin{figure}
\caption{
A plot of the lowest 400 eigenvalues for the non-adiabatic operator in a mixed-valence system.
The parameters are : $\beta\Omega=0.6716$, $\beta\lambda=18.225$, $\beta\D=11.99$, and 
$\beta\epsilon=18.705$. The dot-dashed line is for the case $k=\omega/2\pi$.
}
\end{figure}

\begin{figure}
\caption{ 
Plots of (a) real and (b) imaginary parts of the lowest 30 eigenvalues as a function of 
the coupling constant, $\D$. Except for the coupling constant, all the other parameters are 
set equal to those used in Fig.~1. In Fig. ~2(b), open circles correspond  
to the Rabi frequency $\Omega_{R}=\sqrt{\epsilon^2+4V^2}$. Figure 3(c) shows the three 
dimensional plot of eigenvalues as a function of the coupling constant. 
}
\end{figure}

\begin{figure}
\caption{ 
Comparison between the result of direct numerical propagation and 
spectral propagation. The parameters are chosen as 
$\beta\Omega=1$, $\beta\lambda=15$, $\beta\D=12$, and $\beta\epsilon=3$. 
}
\end{figure}

\begin{figure}
\caption{
Comparison of (a) the dynamics and (b) the spectra in the mixed-valence 
system for three different energy biases. Except for the energy bias, 
all the other parameters are set equal to those used in Fig.~3. Agreements 
between the results of numerical and spectral propagation 
have been checked in these cases.
}
\end{figure}

\begin{figure}
\caption{
(a) Right and (b) left eigenfunctions 
with an eigenvalue $\beta Z=2.6228\pm i 26.394$ 
for a symmetric bias case.$(\beta\epsilon=0)$
All the other parameters are set equal to those used 
in Fig.~3 except for the energy bias. 
Each line corresponds to $\rho_1$(solid), $\rho_2$(dashed),
$u$(dot-dashed), and $v$(dotted), respectively. 
}
\end{figure}

\begin{figure}
\caption{
(a) Right and (b) left eigenfunctions 
with an eigenvalue $\beta Z=2.8057\pm i 26.466$
for an asymmetric bias case.$(\beta\epsilon=5)$
All the other parameters are set equal to those used 
in Fig.~3 except for the energy bias. 
Each line corresponds to $\rho_1$(solid), $\rho_2$(dashed),
$u$(dot-dashed), and $v$(dotted), respectively. 
}
\end{figure}

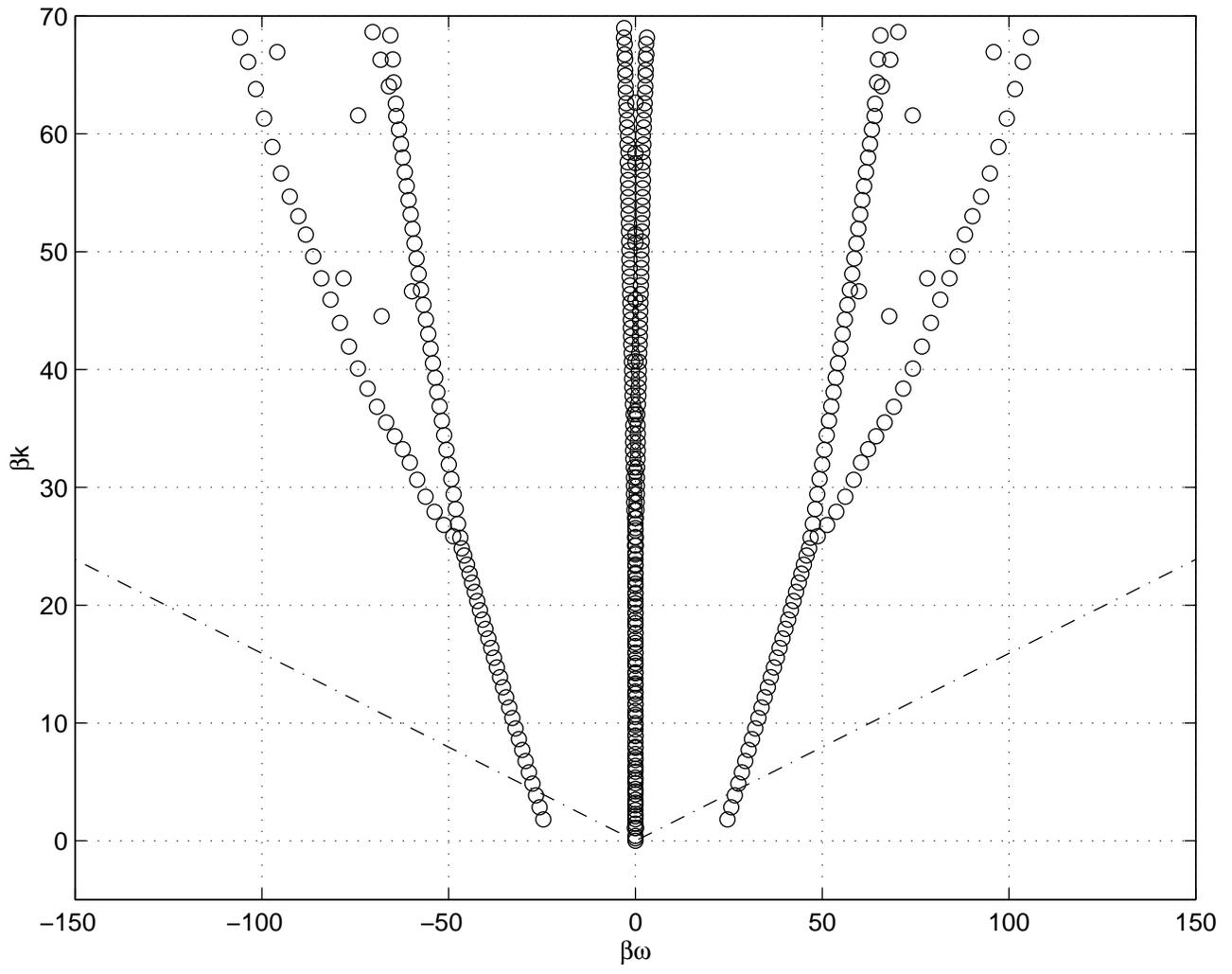
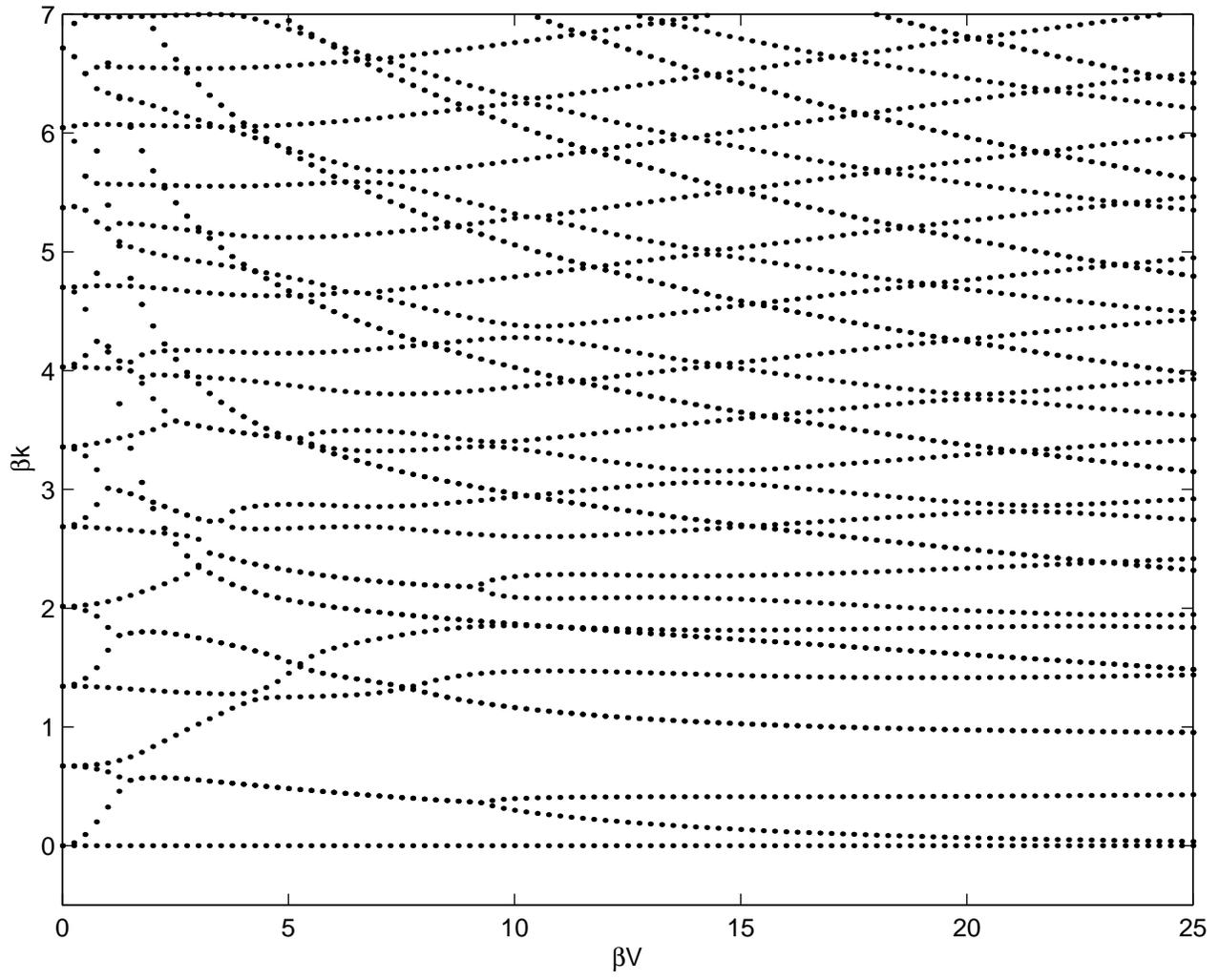
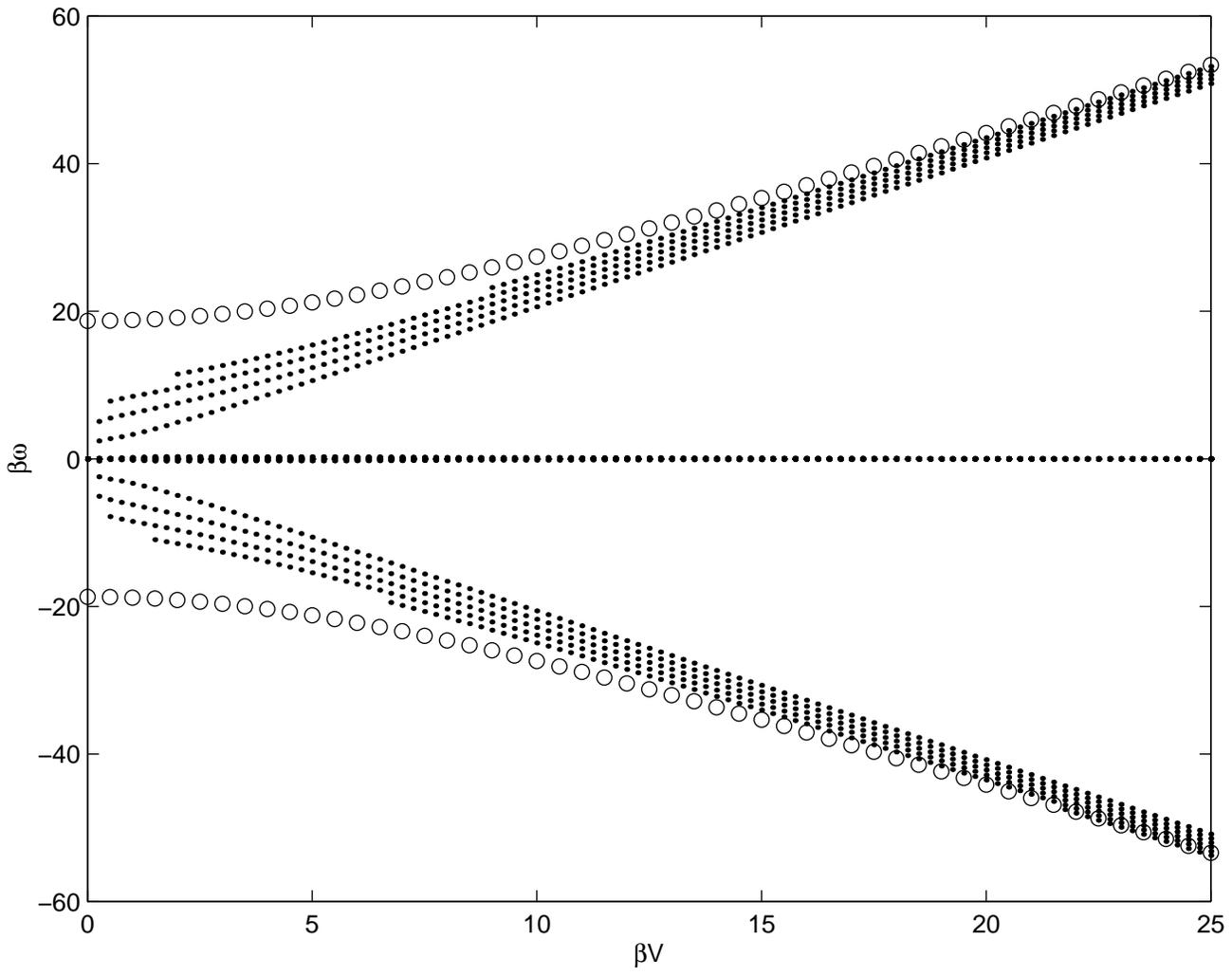
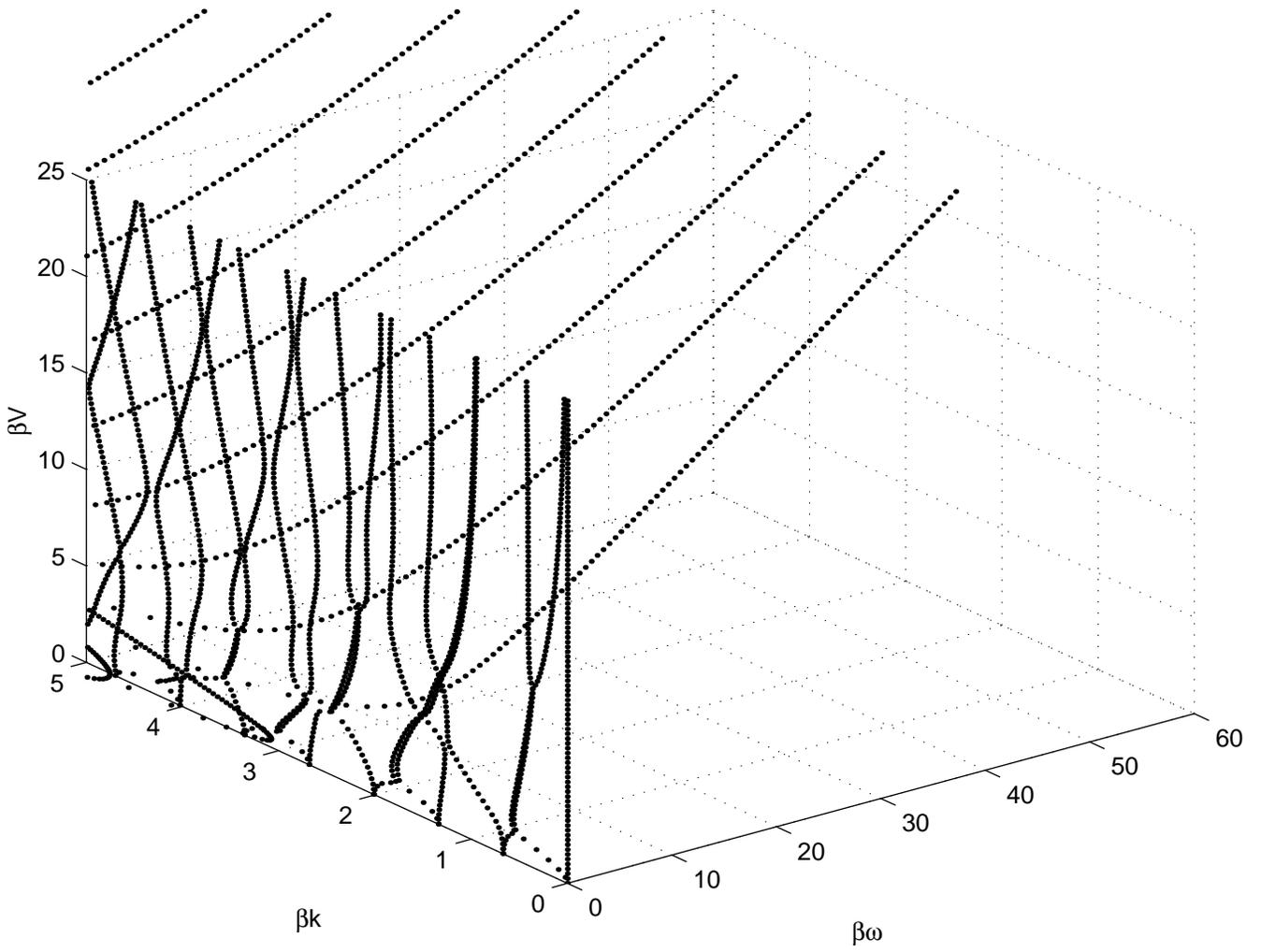
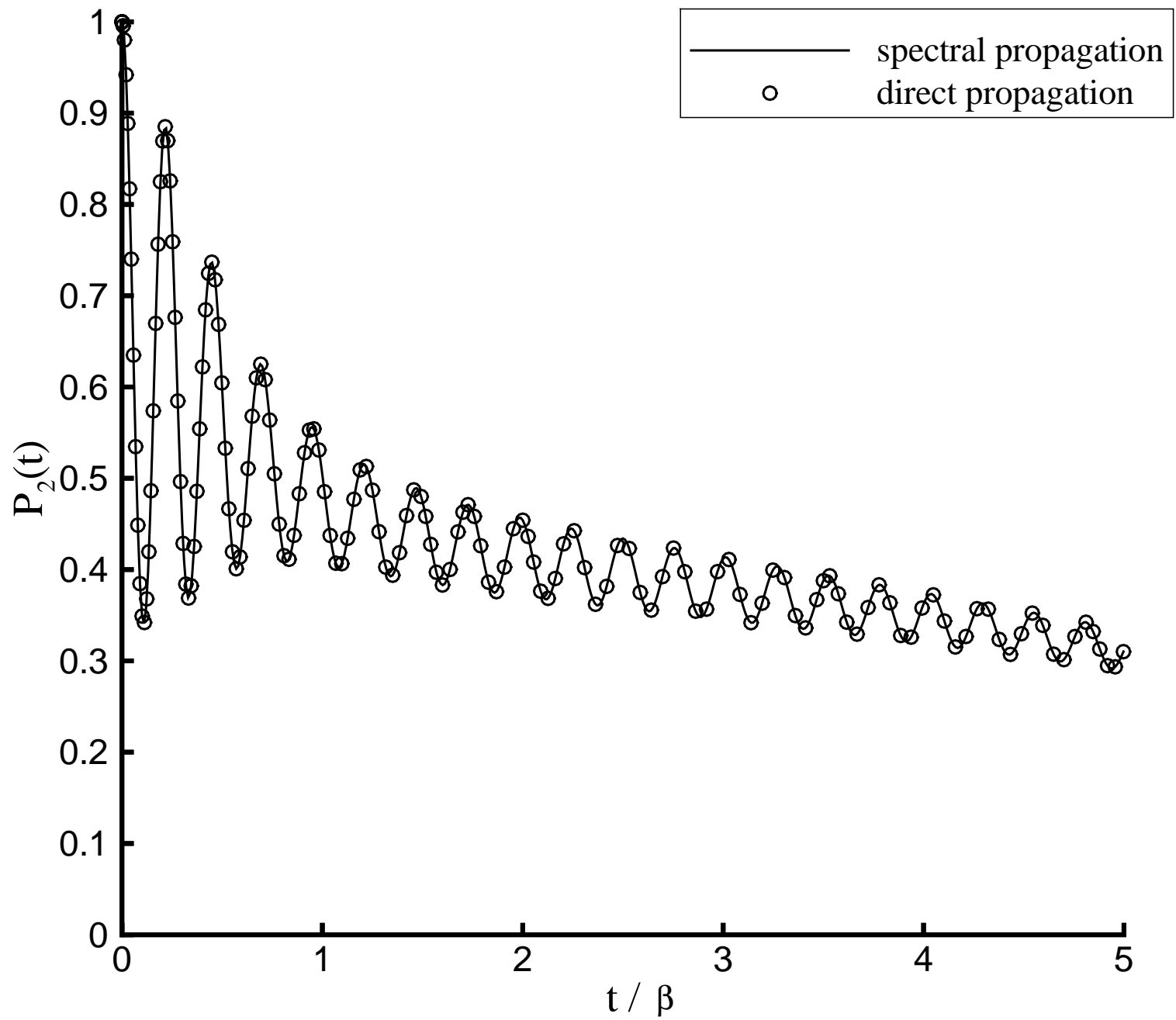
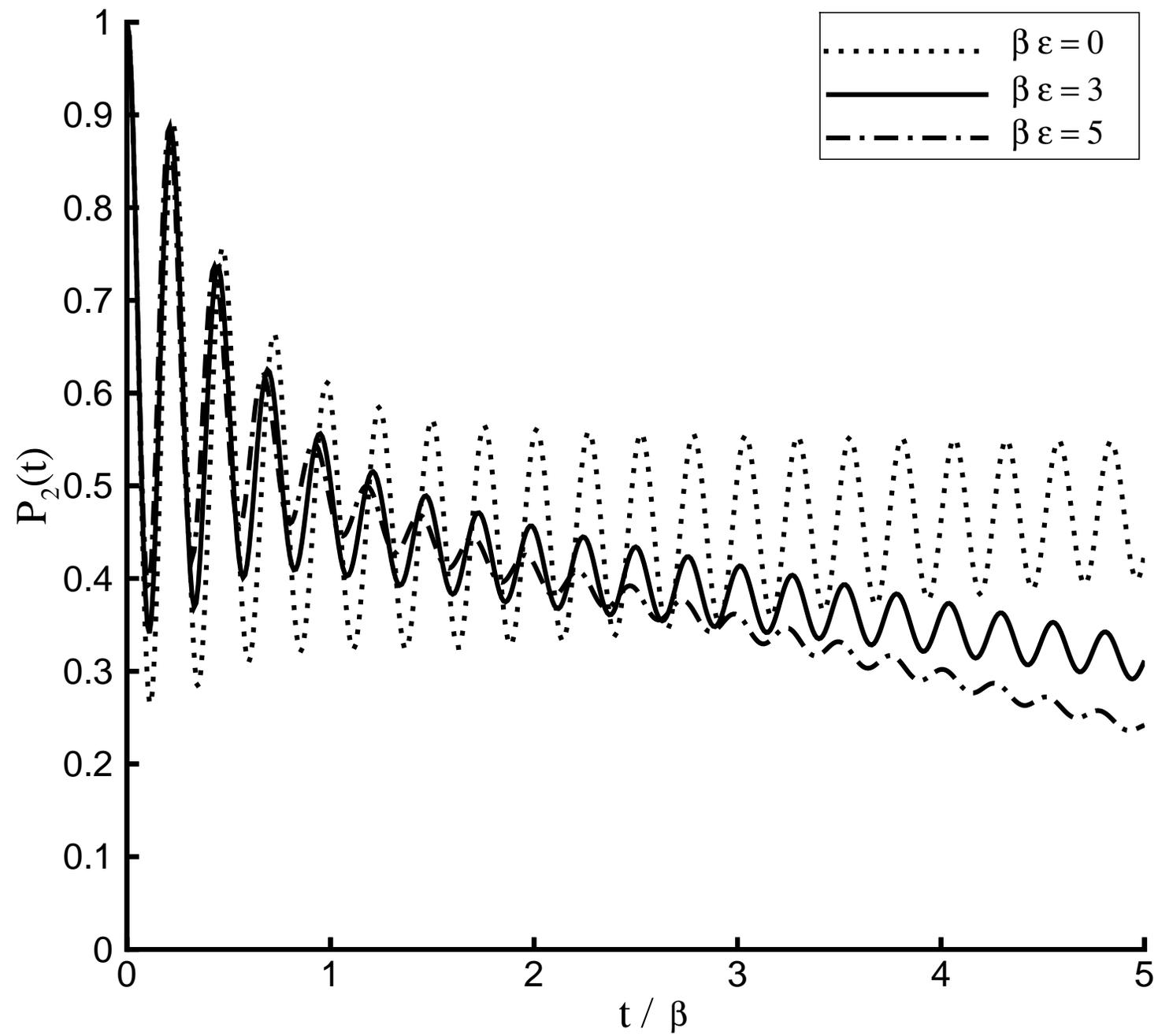
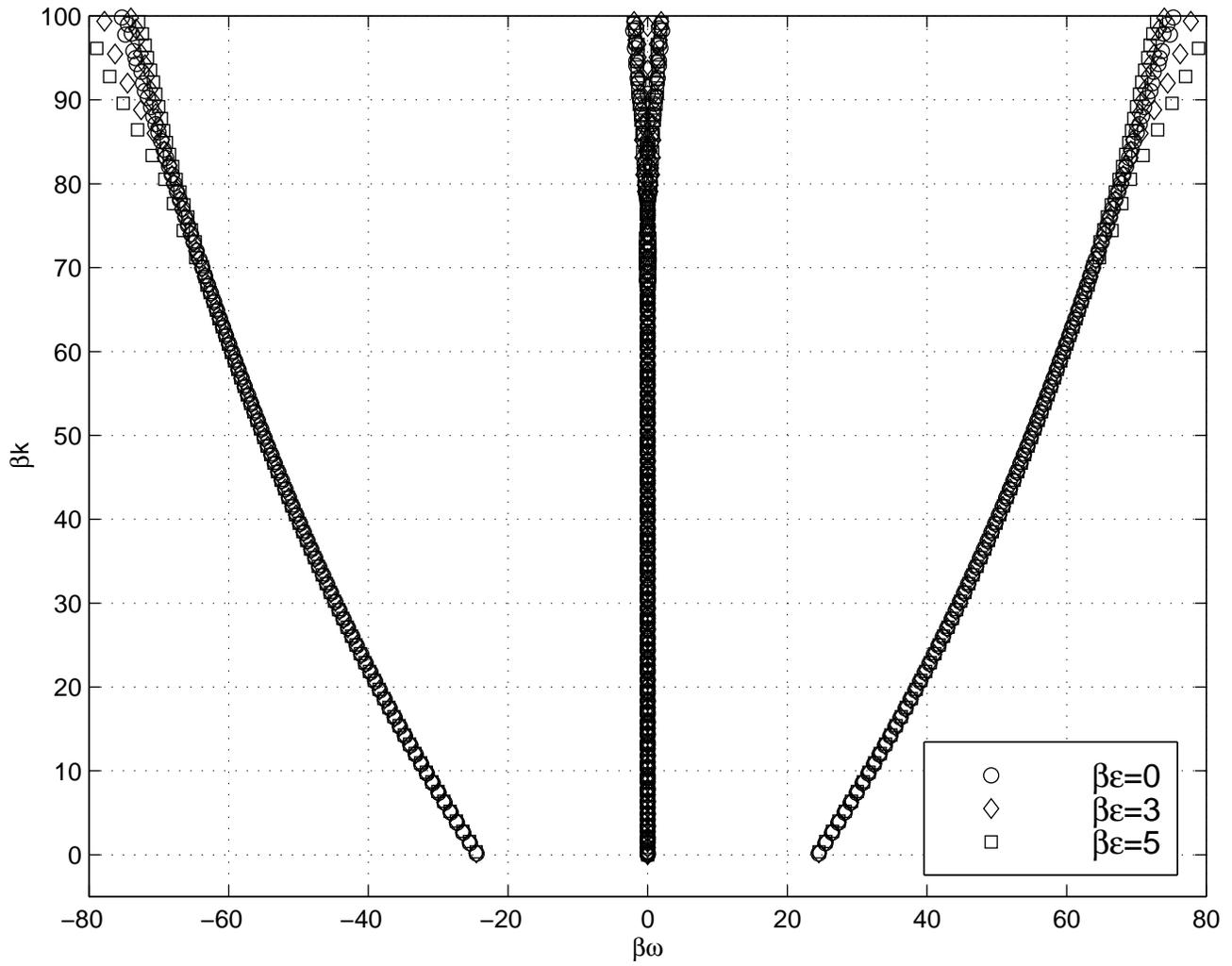
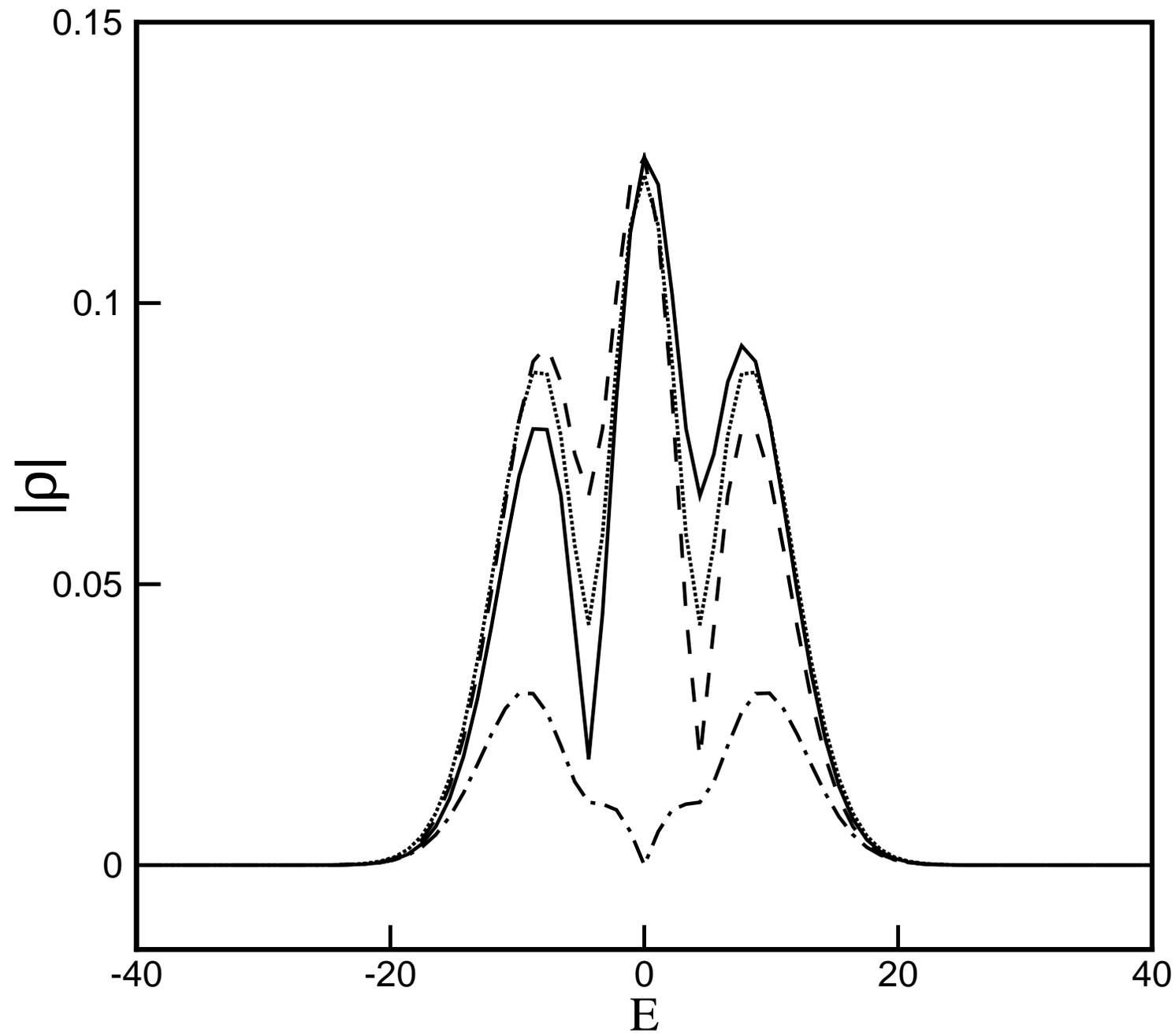
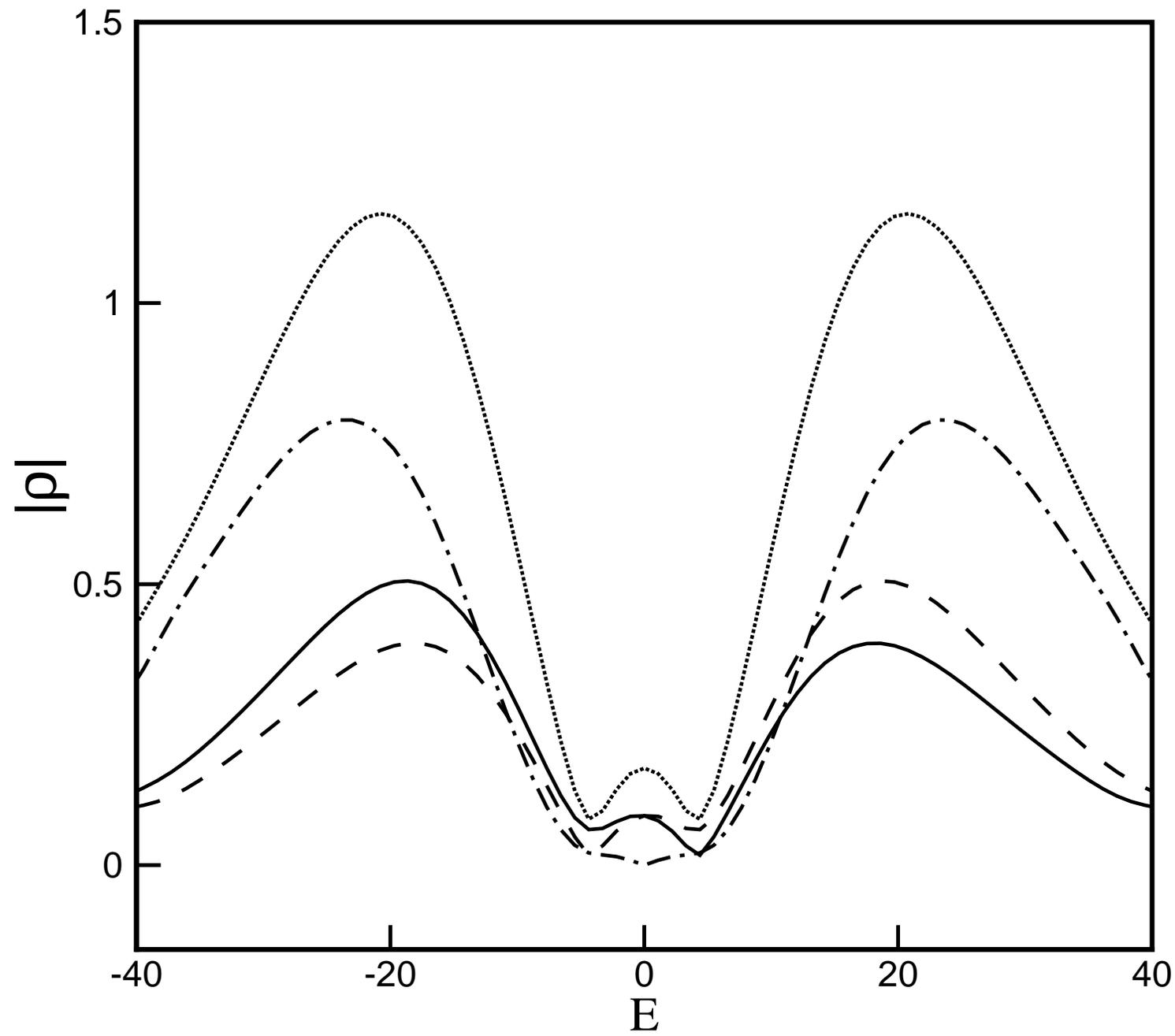
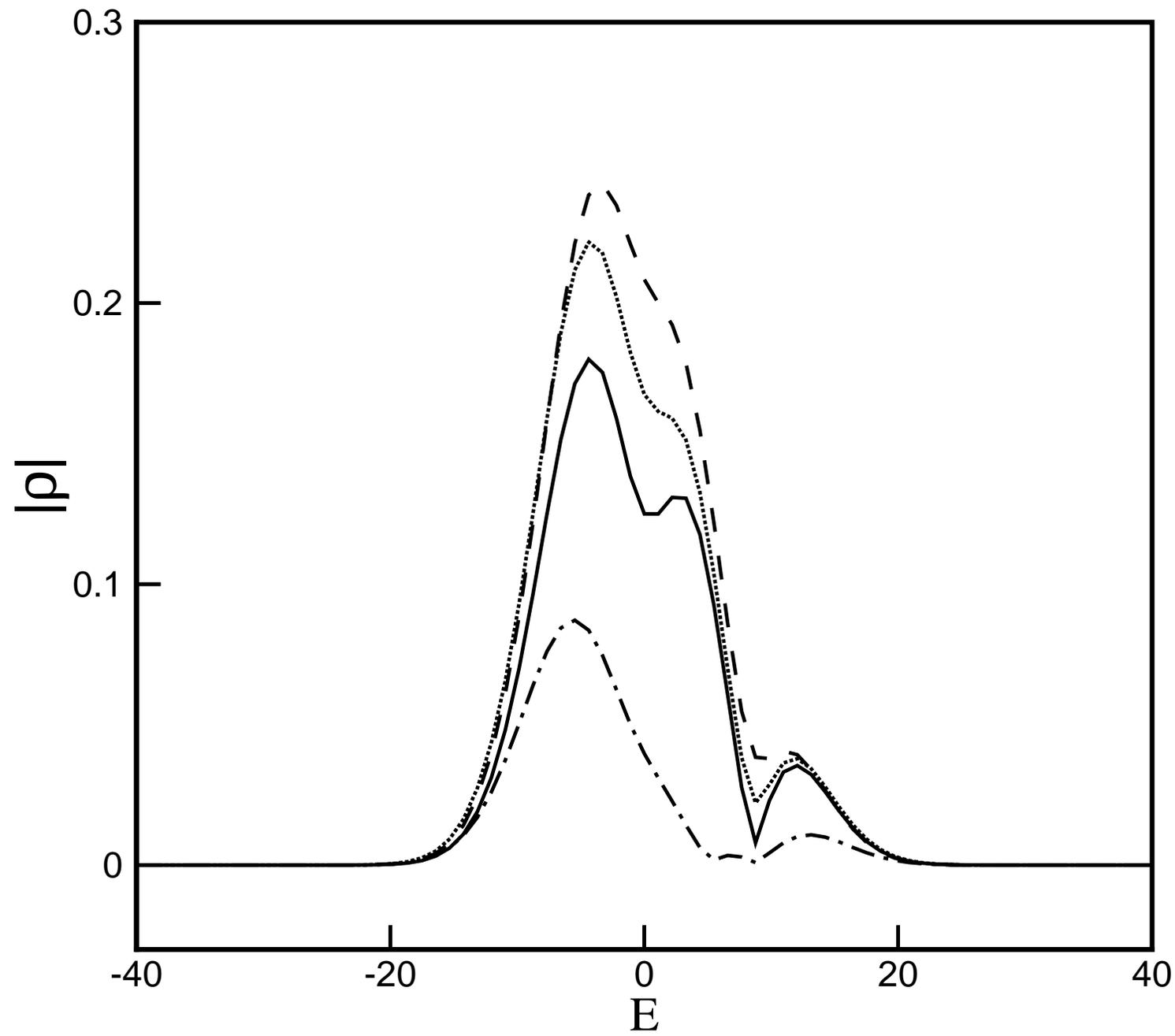
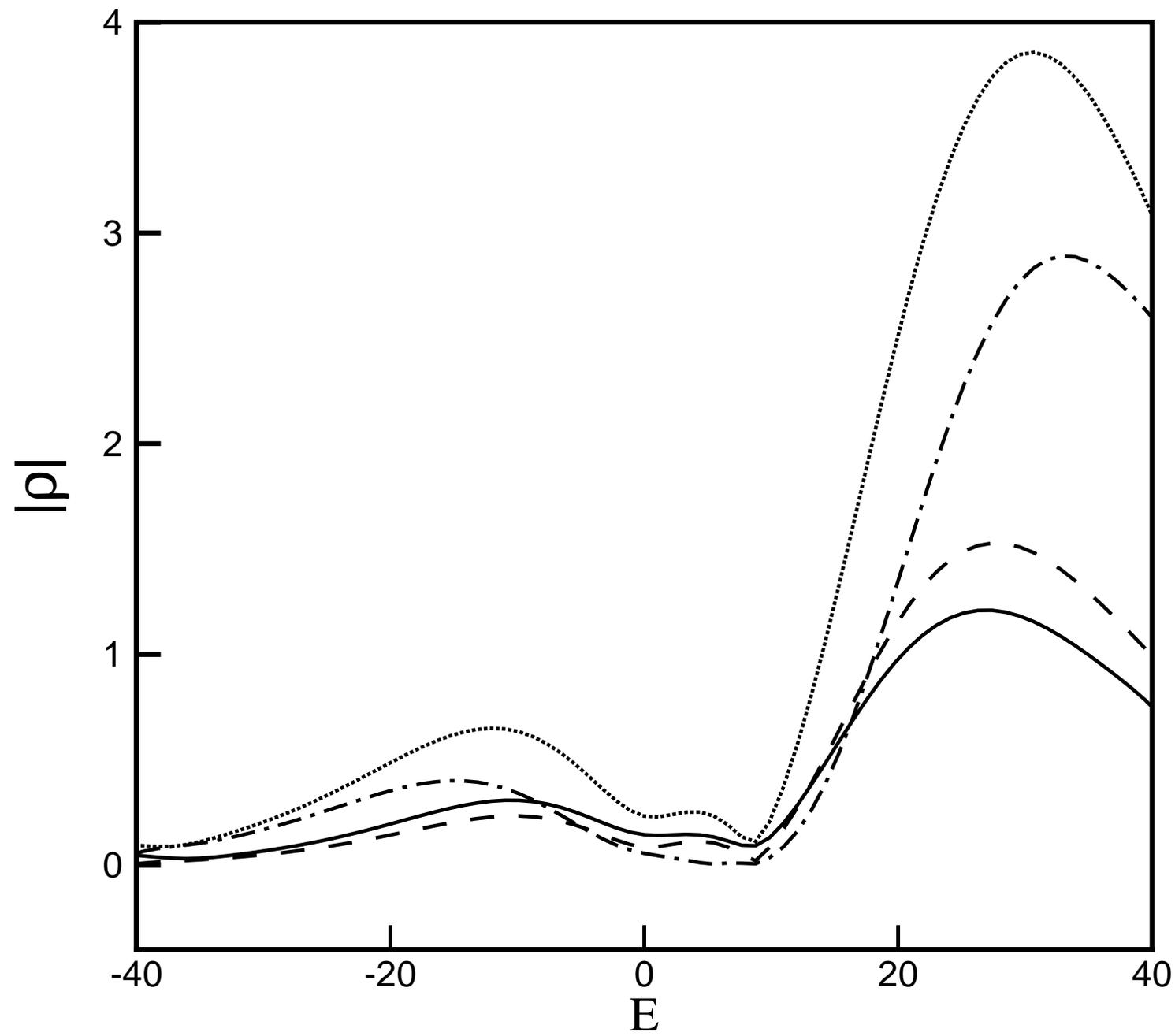
\end{document}